\documentclass[10pt,twocolumn]{revtex4}
\usepackage{graphicx}
\usepackage{amssymb}
\usepackage{amsmath}
\usepackage{bm}
\usepackage{color}

\RequirePackage{pdf14}

\begin{document}

\title{Nonlinear solutions for $\chi^{(2)}$ frequency combs in optical microresonators}

\author{E.\ Podivilov$^{1,2}$, S.\ Smirnov$^2$, I.\ Breunig$^{3,4}$, and B.\ Sturman$^1$}
\affiliation{$^1$Institute of Automation and Electrometry, Russian Academy of Sciences,
630090 Novosibirsk, Russia \\
$^2$Novosibirsk State University, 630090, Novosibirsk, Russia \\
$^3$Laboratory for Optical Systems, Department of Microsystems Engineering - IMTEK, University of Freiburg, Georges-K\"{o}hler-Allee 102, 79110 Freiburg, Germany \\
$^4$Fraunhofer Institute for Physical Measurement Techniques IPM, Heidenhofstra\ss e 8, 79110 Freiburg, Germany
 }

\begin{abstract}
Experimental and theoretical studies of nonlinear frequency combs in $\chi^{(3)}$
optical microresonators attracted tremendous research interest during the last decade
and resulted in prototypes of soliton-based steadily working devices. Realization of similar combs
owing to $\chi^{(2)}$ optical nonlinearity promises new breakthroughs and is a big
scientific challenge. We analyze the main obstacles for realization of the $\chi^{(2)}$
frequency combs in high-$Q$ microresonators and propose two families of steady-state nonlinear 
solutions, including soliton and periodic solutions, for such combs. Despite generic 
periodicity of light fields inside microresonators, the nonlinear solutions can be topologically different 
and relevant to periodic and antiperiodic boundary conditions. The found particular solutions exist
owing to a large difference in the group velocities between the first and second harmonics, 
typical of $\chi^{(2)}$ microresonators, and to the presence of the pump. They have no 
zero-pump counterparts relevant to conservative solitons. Stability issue for the found 
comb solutions remains open and requires further numerical analysis.

\end{abstract}

\date{\today}

\maketitle

\section{Introduction}

Optical microresonators are the subject of numerous experimental and theoretical studies
during the last two decades, as reviewed
in~\cite{Vahala03,MalekiBook09,IlchenkoReviewI,IlchenkoReviewII,Review16}. Ultra-high
(up to $10^{11}$) quality factors of the modes, quasi-discrete modal structure, and
strong enhancement of the light intensity inside owing to the resonant recirculation are
distinctive features of the case. Both $\chi^{(3)}$ and $\chi^{(2)}$ optical materials
(amorphous and crystalline) can be employed. Numerous techniques for engineering of the
modal frequency spectrum and coupling light in and out are well developed. Applications
of microresonators range from sensors of single atoms and molecules to nonlinear optics.
The latter strongly benefit from the possibility to use low-power continuous-wave light
sources.

One of the most spectacular achievements in the nonlinear optics of microresonators
based on $\chi^{(3)}$ materials is generation of broad high-quality frequency combs,
see~\cite{KippNature07,Octave11,KippScience11,Herr12,CombPRL13,KippNP14,Vahala15,Chembo16,Vahala18,KippScience18}
and references therein. It took about one decade to proceed from first observations and
interpretations to the final concept of solitons and its realization on different
platforms~\cite{KippScience18}. According to this concept, the most advanced combs is
manifestation of spatially narrow solitons circulating along the resonator rim with a
constant velocity. The solitons in question are dissipative solitons with a double
balance between ($i$) dispersion broadening and nonlinear narrowing and ($ii$) between
an external pumping and dissipative losses. It turned out that such solitons are stable
against small perturbations. They are essentially different from (and more complicated
than) the so-called conservative solitons existing without pumping and
dissipation~\cite{Akhmediev}.

Application of the soliton concept of frequency combs to microresonators based on
$\chi^{(2)}$ materials represents a big scientific challenge and, simultaneously, a
highly perspective field. It promises employment of lower light powers, of different
spectral ranges, and of new operation regimes. The problem encountered has a certain
basis in the previous studies of conservative $\chi^{(2)}$ spatial and temporal solitons
employing cascading processes of second-harmonic generation and optical parametric
oscillation, see review~\cite{SkryabinReview} and references therein. This basis is,
however, insufficient because of ($i$) incompleteness of knowledge in the field of
conservative solitons, ($ii$) the necessity of transfer to dissipative solitons, and
($iii$) the necessity to incorporate specific features of nonlinear optics of
microresonators. First models for $\chi^{(2)}$ solitons in microresonators were
published recently~\cite{Wabnitz18,Skryabin19,Skryabin19A}. Attempts to explore $\chi^{(2)}$
frequency combs regardless of solitons are reported in~\cite{Att1,Att2,Att3,Att4}. Also, the 
so-called simulton-solitons are reported recently for meter-scale parametric oscillators~\cite{Fejer18}. 
The area remains, in essence, largely unexplored.

The following features specific for the $\chi^{(2)}$ frequency combs have to be indicated: \\
-- The first and second harmonics (FH and SH) in $\chi^{(2)}$ materials are not
phase-matched in the general case, so that their nonlinear coupling is negligible except
for special cases. The quasi-phase-matching employing proper radial poling of
microresonators~\cite{RadialPoling1,RadialPoling2} has to be used to overcome this
problem. Properties of the radial poling have to be special for each operation regime. \\
-- The group velocities of FH and SH are generally different, so that the corresponding
wave envelopes separate from each other unless the nonlinear coupling compensates for
this separation. The effect of the group velocity difference is typically very strong
and dominating over the effects of frequency dispersion. Very little is known about
dispersionless soliton regimes. \\
-- In the cases when the quasi-phase-matching is combined with a zero group velocity
difference and the effects of frequency dispersion are important, the signs of the
frequency dispersion are typically opposite for FH and SH. This imposes additional
restrictions on the soliton regimes. \\
-- There are two principal cases for generation of $\chi^{(2)}$ frequency combs --
pumping into FH and SH modes. In the first case, primarily pumped monochromatic FH
excites a SH, which can be unstable against generation of side FHs. Here, the spatially
uniform background states for FH and SH exist and represent an important ingredient of
soliton physics. In the second case of pumped SH, the spatially uniform FH and SH
backgrounds are not necessarily exist leading to specific comb generation regimes.

In this paper, we focus on the case of SH pumping and dominating effect of the group
velocity difference. At the same time, our basic equations incorporate the effects of
frequency dispersion. As in the previous approaches to comb modeling, we assume that
each resonator mode can be characterized by a single modal number. This assumption can
be implemented by proper shaping of the resonator rim. It strongly simplifies the
considerations.

\section{Basic equations}

\begin{figure}[h]
\centering
\includegraphics[width=8cm]{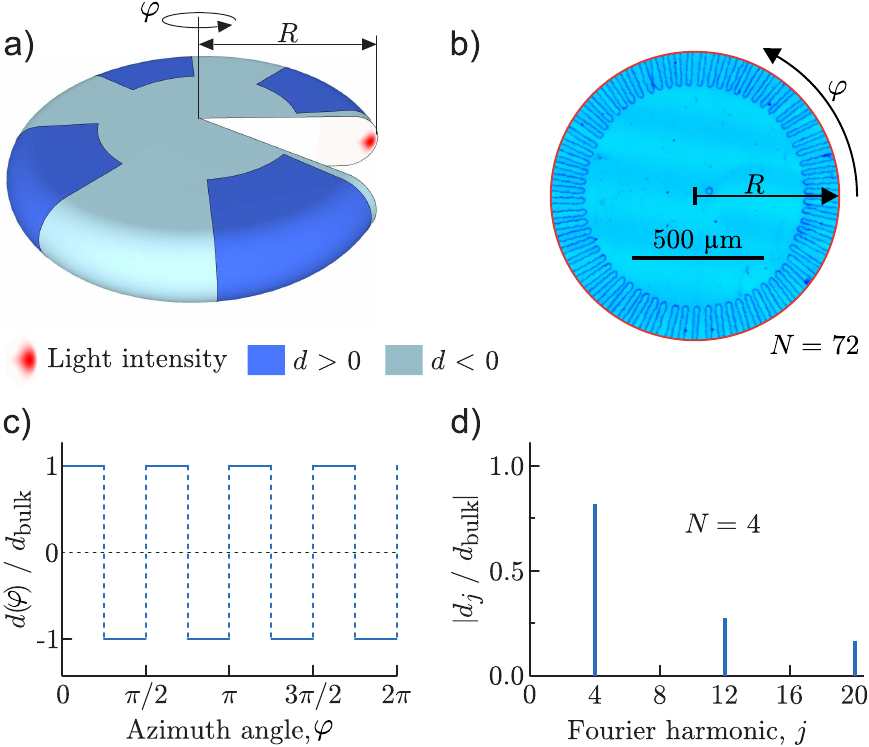}
\caption{Resonator geometry and radial poling. a) Resonator shape and cross section, $\varphi$ is 
the azimuth angle, the major radius $R$ is different from the rim curvature radius. The red 
spot shows localization of light, and different colors indicate the radial poling. b) Micrograph 
of radially poled lithium niobate based resonator with the number of domain periods $N = 72$. 
c) Schematic of the azimuth dependence of the nonlinear coefficient $d(\varphi)$ for perfectly periodic
poling with $N = 4$. d) First harmonics of the Fourier spectrum $|d_j|$ for~c).}\label{Geometry}
\end{figure}

The light modes (whispering gallery modes) can be viewed as quasi-one-dimensional waves,
with wavevectors $k_j = j/R$ and very large ($\sim 10^4$) integer azimuth mode number
$j$, propagating along the rim of an axisymmetric resonator with big radius $R$, see
Fig.~\ref{Geometry}a.
The modal light fields depend on the azimuth angle $\varphi$ as $\exp(ij\varphi)$.
Details of localization of light fields near the rim for the chosen modes are of minor importance 
for this study. The modal frequencies are discrete and given by $\omega_j = k_jc/n_j$, where $c$ 
is speed of light and $n_j$ is the modal refractive index. Frequency $\omega_j$ corresponds to 
the vacuum wavelength $\lambda_j = 2\pi c/\omega_j$, and the modal refractive index can be treated 
as a smooth function $n(\lambda)$ to characterize the modal group velocity and dispersion. 
For $\chi^{(2)}$ resonators, radius $R$ is typically of the order of $1$~mm. Here, the smooth 
function $n(\lambda)$ is close to the function $n_0(\lambda)$ characterizing the bulk material. 
Small corrections $n - n_0$ are size and shape dependent~\cite{Gorodetsky1,Gorodetsky2,We19} and 
can be taken into account for application to particular resonators. Generally, two polarization 
types of modes, with different dependences $n(\lambda)$, exist in microresonators.

In the case of azimuth symmetry, the phase-matching condition for SH generation is $2\omega_j 
= \omega_{2j}$. It is valid also for parametric generation of mode $j$ by mode with the azimuth 
number $2j$. Except for special cases, this phase-matching condition cannot be fulfilled~\cite{Nikogosyan}. 
Moreover, the frequency distance $2\omega_j - \omega_{2j}$, being much smaller than $\omega_j$, 
is usually much larger than the intermodal distance $\delta \omega = c/nR$. 

Periodic radial poling is in use to provide phase-matching in a designated spectral 
range~\cite{RadialPoling1,RadialPoling2}. It is illustrated by Figs.~\ref{Geometry}b,c,d. 
Ideally, the radial poling does not influence the linear optical properties and results  
in a strictly periodic in $\varphi$ alternation of the sign of the quadratic susceptibility coefficient 
$d$, as indicated in Figs.~\ref{Geometry}c. If $N$ is the number of alternation periods, the azimuth 
dependence $d(\varphi)$ is given by the Fourier series $d = d_0 + d_1\exp({\rm i}N\varphi) + 
d_3\exp(3{\rm i}N\varphi) + \ldots$, where $|d_1|$ can be comparable with the bulk value of 
$d$~\cite{RadialPoling1}. The Fourier spectrum of $d(\varphi)$ for the perfect 
poling is illustrated by Fig.~\ref{Geometry}d. The higher peaks are typically unimportant 
for phase matching. The phase-matching condition for SH generation reads now $2\omega_j = 
\omega_{2j \pm N}$, where the sign "plus" is relevant to the most typical case of decreasing 
$n(\lambda)$. By choosing $N$, it can be fulfilled with a good accuracy (within the intermodal 
distance) for any $j$. Modification of the phase-matching conditions for generation of sum 
and difference frequencies can be made similarly.

Now, we write down generic nonlinear equations for complex modal amplitudes $F_{j}$ and
$S_{l}$ relevant to the FH and SH frequency domains and the phase-matching condition
$2\omega_j = \omega_{2j + N}$. Assuming pumping into a SH mode with modal number $l_0$,
we have
\begin{eqnarray}\label{Initial}
\hspace*{-4mm} &{\rm i}& \hspace*{-2mm}\dot{F}_{\hspace*{-0.3mm} j} \hspace*{-0.5mm} -
\hspace*{-0.5mm} (\omega_{j} \hspace*{-0.5mm} - \hspace*{-0.5mm}
{\rm i}\gamma_{j})F_{\hspace*{-0.3mm} j} \hspace*{-0.5mm} = \hspace*{-0.5mm} 2\mu
\hspace*{-0.5mm} \sum\limits_{l,j'} S_{l}F^*_{\hspace*{-0.3mm} j'} \, \delta_{l
\hspace*{-0.3mm} - \hspace*{-0.3mm} j \hspace*{-0.3mm} -
\hspace*{-0.3mm} j' \hspace*{-0.3mm} - \hspace*{-0.3mm} N}  \\
\hspace*{-4mm} &{\rm i}&\hspace*{-2mm}\dot{S}_{l} \hspace*{-0.5mm} - \hspace*{-0.5mm}
(\omega_{l} \hspace*{-0.5mm} - \hspace*{-0.5mm} {\rm i}\gamma_{l})S_{l} \hspace*{-0.5mm} =
\hspace*{-0.6mm} \mu \hspace*{-0.5mm} \sum\limits_{j,j'} \hspace*{-0.5mm}
F_{\hspace*{-0.3mm} j}F_{\hspace*{-0.3mm} j'} \delta_{l \hspace*{-0.3mm} -
\hspace*{-0.3mm} j \hspace*{-0.3mm} - \hspace*{-0.3mm} j' \hspace*{-0.3mm} -
\hspace*{-0.3mm} N} \hspace*{-0.5mm} + \hspace*{-0.5mm} {\rm i}h\delta_{l \hspace*{-0.3mm} -
\hspace*{-0.3mm} l_0}e^{-{\rm i}\omega_pt} . \nonumber
\end{eqnarray}
Here the dot indicates differentiation in time $t$, $\gamma_{j,l}$ are the modal decay
constants, such that the quality factor is $Q_j = \omega_j/2\gamma_j \gg 1$, $\mu$ is a
coupling coefficient incorporating the relevant susceptibility coefficient and modal
overlaps~\cite{IlchenkoReviewI,Generic}, $\omega_p$ is the pump frequency, $h$ is a
variable parameter characterizing pump strength, and the asterisk indicates complex
conjugation. The amplitudes are normalized such that $\omega_j|F_j|^2$ and
$\omega_l|S_l|^2$ are the modal energies. Without loss of generality, $\mu$ and $h$ can
be treated as real positive quantities. The modes $F$ and $S$ can be polarizationally
the same or different. The case $N = 0$ corresponds to phase matching of differently
polarized modes without radial poling~\cite{NaturalPM}. Transfer to the case of FH
pumping is evident. Note that the presence of a factor of $2$ in Eqs.~(\ref{Initial})
reflects the Hamiltonian nature of $\chi^{(2)}$ interactions in the absence of
dissipation.

Further simplifications involve an assumption of relative narrowness of the FH and SH
spectra. Let the numbers $j$ and $l$ be close to $j_0$ and $l_0$, respectively, so that
the deviations $\delta j = j - j_0$ and $\delta l = l - l_0$ are much smaller than $j_0$
and $l_0$ in absolute values. Then we can employ expansions
\begin{eqnarray}\label{Expansions}
\omega_{j} &=& \omega_{j_0} + \nu_1 (j - j_0) + \alpha_1 (j - j_0)^2 \\
\omega_{l} &=& \omega_{l_0} + \nu_2 (l - l_0) + \alpha_2 (l - l_0)^2 \;, \nonumber
\end{eqnarray}
where $\nu = v/R$, $\alpha = v'/2R^2$, $v = d\omega_k/dk$ is the group velocity, $v' =
dv/dk = d^2\omega_k/dk^2$ is the dispersion parameter, and indices $1$ and $2$ refer to
$k_1 = j_0/R$ and $k_2 = l_0/R$. Parameters $v$ and $v'$ are fully determined by the
dependence $n(\lambda)$.
\begin{figure}[h]
\centering
\includegraphics[width=8.4cm]{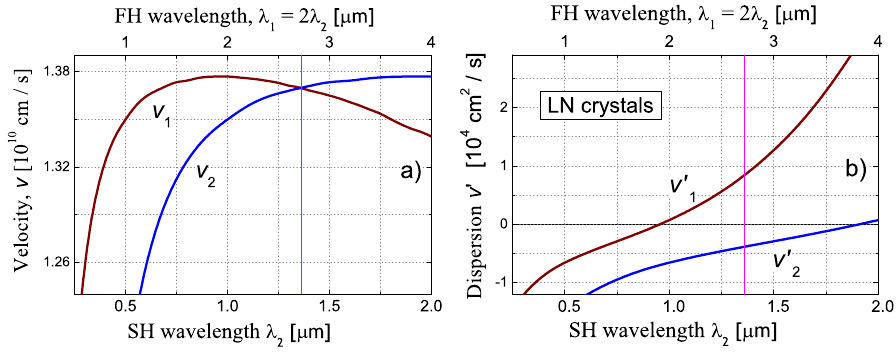}
\caption{Wavelength dependences of $v_{1,2}$ (a) and $v'_{1,2}$ (b) for LiNbO$_3$
crystals with the geometric dispersion neglected. The vertical line corresponds to the
case $v_1 = v_2$, where $v'_1v'_2 < 0$.}\label{LN-curves}
\end{figure}
Figure~\ref{LN-curves} shows representative wavelength dependences of $v_{1,2}$ and $v'_{1,2}$ 
for LiNbO$_3$ crystals relevant to the bulk value $n_0(\lambda)$ and extraordinary
polarization~\cite{LN1,LN2}. They show that the velocity difference $v_1 - v_2$ ranges
from $0$ to $\approx 10^{-2}c$, it can be positive and negative. The dispersion
parameters can be of the same and opposite signs, they turn to zero at certain
wavelengths. Influence of the resonator shape and size is minor for $R \sim 1$~mm. The
mentioned features are crucial for the subsequent analysis of nonlinear comb solutions.

Using Eqs.~(\ref{Initial}) and (\ref{Expansions}), we can obtain equations for slowly
varying amplitudes $F(\varphi,t)$ and $S(\varphi,t)$. The structure of these equations
and the subsequent results depend, however, on whether the number $l_0 - N$ in
Eqs.~(\ref{Initial}) is even or odd. This leads us to an important notion of periodic
and antiperiodic solutions for $F(\varphi,t)$ and $S(\varphi,t)$. This notion is not
applicable to the $\chi^{(3)}$ case.

\section{Periodic and antiperiodic solutions}

Let the integer $l_0 - N$ be {\em even}. Then $\omega_{l_0}$ is close to $2\omega_{j_0}$
with $j_0 = (l_0 - N)/2$, so that the frequency mismatch $\Delta_0 = 2\omega_{j_0} -
\omega_{l_0}$ is smaller in the absolute value than the intermodal distance $\delta
\omega = c/nR$. It is convenient here to use the discrete deviations $\delta j = j -
j_0$ and $\delta l = l - l_0$. The slowly varying amplitudes $F(\varphi,t)$ and
$S(\varphi,t)$ are linked to $F_{\delta j}$ and $S_{\delta l}$ by the relations of
discrete Fourier transform
\begin{eqnarray}\label{SlowlyVarying}
F &=& e^{{\rm i}(\omega_{j_0} - \Delta_1)t} \sum\limits_{\delta j} F_{j_0 + \delta j}
e^{{\rm i}\delta j\varphi} \\
S &=& e^{{\rm i}(\omega_{l_0} - \Delta_2)t} \sum\limits_{\delta l} S_{l_0 + \delta l}
e^{{\rm i}\delta l\varphi} \,,  \nonumber
\end{eqnarray}
where $\Delta_1 = (2\omega_{j_0} - \omega_p)/2$ and $\Delta_2 = \Delta_p = \omega_{l_0}
- \omega_p$. To get equations for $F$ and $S$, it is necessary to multiply the first and
second of Eqs.~(\ref{Initial}) by $\exp({\rm i}\delta j \varphi)$ and $\exp({\rm i}\delta
l\varphi)$, respectively, take sums in $\delta j$ and $\delta l$, and employ
Eqs.~(\ref{Expansions},\,\ref{SlowlyVarying}) and the Kronecker symbols. Finally, we
obtain
\begin{equation}\label{Initial2}
\hat{L}_1 F = 2\mu SF^*, \qquad \hat{L}_2 S = \mu F^2 + {\rm i}h \;,
\end{equation}
where the linear differential operators $\hat{L}_{1,2}$ are given by
\begin{equation}\label{L}
\hat{L}_{1,2} = {\rm i}(\partial_t + \nu_{1,2}\partial_{\varphi}) +
\alpha_{1,2}\partial^2_{\varphi} - \Omega_{1,2}
\end{equation}
with $\Omega_{1,2} = \Delta_{1,2} - {\rm i}\gamma_{1,2}$. The $\nu$ and $\alpha$ terms
correspond to the effects of drift and dispersion of wave envelopes, respectively. These effects 
are generally substantially different for FH and SH amplitudes. For simplicity, we have
neglected the wavelength dependences of small damping coefficients $\gamma_{1,2}$.
According to the definition (\ref{SlowlyVarying}), the amplitudes $F(\varphi,t)$,
$S(\varphi,t)$ and their derivatives obey $2\pi$-periodic boundary conditions. Detunings
$\Delta_1$ and $\Delta_2$ have different meaning. The pump detuning $\Delta_2$ can be
easily tuned in experiment, while detuning $\Delta_1$ is greatly determined by the
material and resonator properties, it can be affected, e.g., by temperature tuning.

Remarkably, Eqs.~(\ref{Initial2}) possess spatially uniform steady-state solutions
$\bar{F}$, $\bar{S}$; they are important for analysis of soliton solutions. Below the
threshold of parametric instability, $h < h_{\rm th} = |\Omega_1\Omega_2|/2\mu$, we have
a trivial solution $\bar{F} = 0$ and $\bar{S} = -{\rm i}h/\Omega_2$. For $\bar{F} \neq 0$, we
obtain $|\bar{S}| = h_{\rm th}/|\Omega_2| = |\Omega_1|/2\mu$ and
\begin{equation}\label{Background}
\frac{2\mu^2|\bar{F}|_{\pm}^2}{|\Omega_1\Omega_2|} = \cos \Phi \pm \sqrt{\eta^2 -
\sin^2\Phi} \; ,
\end{equation}
where $\eta = h/h_{\rm th} = 2\mu h/|\Omega_1\Omega_2|$ is the normalized 
pump strength and $\Phi = {\rm arg}(\Omega_1\Omega_2)$.
\begin{figure}[h]
\centering
\includegraphics[width=7.8cm,height=5.3cm]{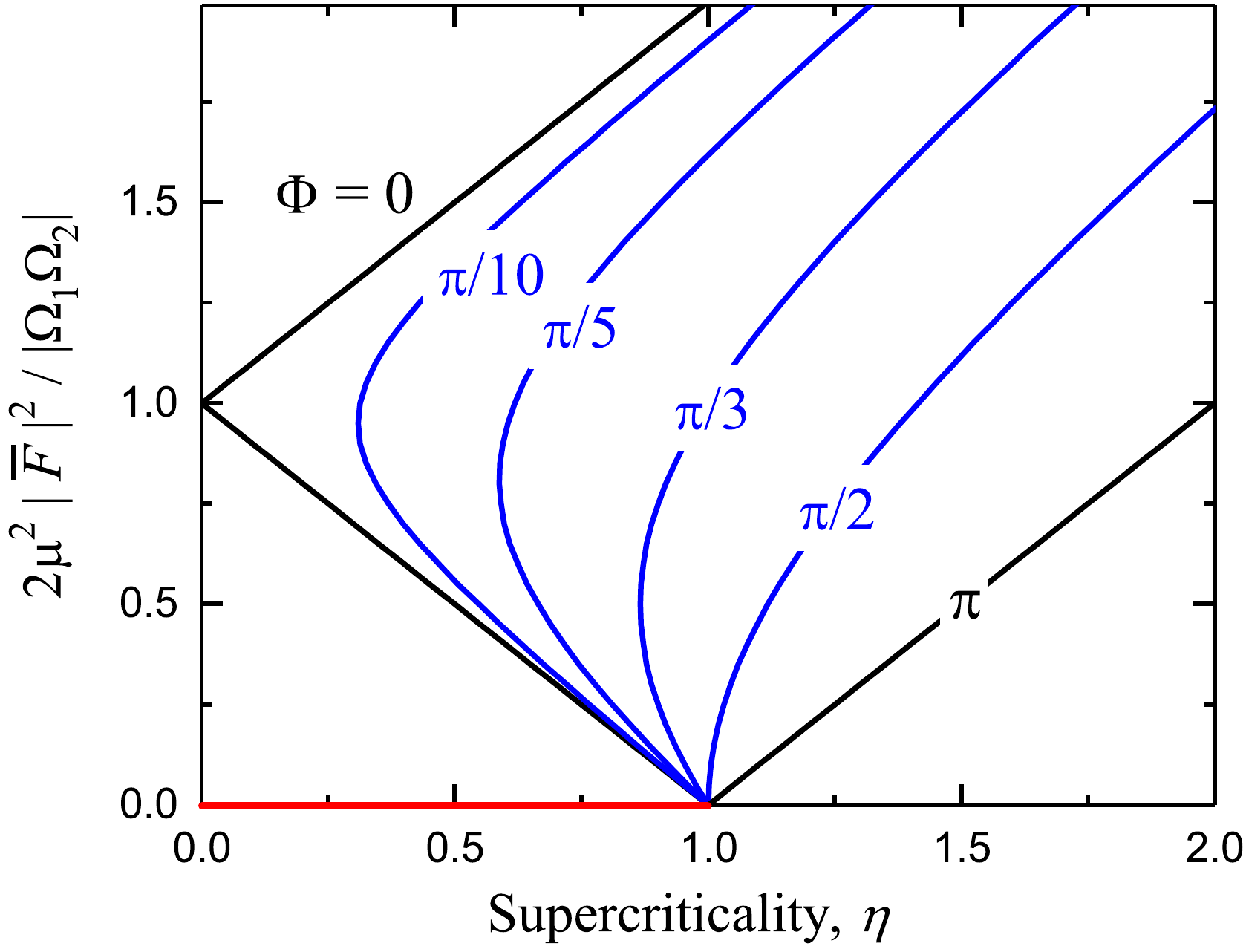}
\caption{Dependences $|\bar{F}_{\pm}|^2(\eta)$ for different values of $\Phi = {\rm
arg}(\Omega_1\Omega_2)$. The $\pm$ branches correspond to positive/negative slopes.
The two-branch black curve with $\Phi = 0$ corresponds to the limiting case $\gamma_{1,2}
\to 0$ at $\Delta_1\Delta_2 > 0$, while the black line $\Phi = \pi$ is relevant to the 
liming case $\gamma_{1,2} \to 0$ at $\Delta_1\Delta_2 < 0$ and to the case $\Delta_1 = 
\Delta_2 = 0$. The red segment shows the branch $\bar{F} = 0$ for $\bar{S} \neq 0$. }\label{Back}
\end{figure}
Figure~\ref{Back} illustrates the dependences $|F_{\pm}|^2(\eta,\Phi) \neq 0$. For $0 \leq|\Phi|
< \pi/2$, there are two branches, while for $\pi/2 \leq |\Phi| \leq \pi$ there is only
single branch. For $|\Phi| \ll 1$, the minimum value of $\eta$ is $\eta_{\min} \simeq
|\Phi|$. The branches $\bar{F}_-(\eta)$ are expected to be unstable against spatially
uniform temporal perturbations, see also below. This means the presence of bistable
background solutions for $\eta < 1$, $|\Phi| < \pi/2$ and, possibly, hysteresis when
adiabatically increasing and decreasing~$\eta$.

Let now the integer $l_0 - N$ be {\em odd}. It cannot be equal to a double integer
$2j_0$. In other words, the $l_0$ mode is coupled not to a single FH mode but to two FH
modes. This means, in particular, that solutions with spatially uniform intensities
$|F|^2$ and $|S|^2$ are not possible. The phase matching conditions say here that $l_0 -
N = 2j_0 + 1$ and $\omega_{l_0}$ is close to $\omega_{j_0} + \omega_{j_0 + 1}$. Two
primary FHs $j_0$ and $j_0 + 1$ not only influence the pumped SH $l_0 = 2j_0 + 1$, but
force inevitably SHs with $\l_0 \pm 1$ etc. leading inevitably to broadening of both
spectra.

Our nearest goal is to introduce slowly varying amplitudes $F(\varphi,t)$ and
$S(\varphi,t)$ obeying again Eqs.~(\ref{Initial2},\,\ref{L}). This can be done with
expressions $F = \tilde{F}\exp(-{\rm i}\varphi/2)$,
\begin{eqnarray}\label{FS-Antiperiodic}
\tilde{F} &=& e^{{\rm i}(\omega_{j_0} - \Delta_1 + \delta\omega_1/2)t} \sum\limits_{\delta j}
F_{j_0 +
\delta j} \; e^{{\rm i}\delta j\varphi} \nonumber \\
S &=& e^{{\rm i}(\omega_{l_0} - \Delta_2)t} \sum\limits_{\delta l} S_{l_0 + \delta l}
\;e^{{\rm i}\delta l \varphi},
\end{eqnarray}
where $\Delta_1 = (\omega_{j_0} + \omega_{j_0 + 1} - \omega_p)/2$ and, as earlier,
$\Delta_2 = \Delta_p = \omega_{l_0} - \omega_p$. Here, the most remarkable feature is
the presence of factor $\exp(-{\rm i}\varphi/2)$ in the definition of $F$. Owing to this
factor, the amplitude $F(\varphi,t)$ is $2\pi$-{\em antiperiodic}, whereas the amplitude
$S(\varphi,t)$ is $2\pi$-periodic.

Note that the FH amplitude $\tilde{F}(\varphi,t)$, which differs from $F(\varphi,t)$ by
a factor of $\exp({\rm i}\varphi/2)$, remains $2\pi$-periodic. Employment of the pair of
amplitudes $\tilde{F},\,S$ leads us to the set
\begin{equation}\label{Initial3}
\hat{L}'_1 \tilde{F} = 2\mu S\tilde{F}^*e^{{\rm i}\varphi}, \qquad \hat{L}_2 S = \mu
\tilde{F}^2e^{-{\rm i}\varphi} + {\rm i}h \;,
\end{equation}
where operator $\hat{L}'_1$ is different from $\hat{L}_1$, see Eq.~(\ref{L}), by the
replacement $\Delta_1 \to \Delta'_1 = \Delta_1 - \delta \omega_1/2$. In contrast to
set~(\ref{Initial2}), it includes $2\pi$-periodic factors $\exp(\pm i\varphi)$ in the
right-hand sides. This is inconvenient for analytical treatments but useful for
numerical calculations.

Modification of the boundary conditions for Eqs.~(\ref{Initial2},\,\ref{L}) in the case
of odd number $l_0 - N$ extends the range of physical solutions. In particular, only
solitons with antiperiodic boundary conditions for $F$ can be available, see below. It
is also clear that this modification is not compatible with the presence of spatially
uniform solutions. In the case of FH pumping, the antiperiodic solutions are absent.

\section{Steady-state solutions}
\subsection{General properties}

Steady-state nonlinear solutions, where the FH and SH envelopes propagate with the same
velocity, are of our prime interest. For such a solution we have $F(\varphi,t) =
F(\tilde{\varphi})$ and $S(\varphi,t) = S(\tilde{\varphi})$, where $\tilde{\varphi} =
\varphi - \nu_0t$ is the moving frame coordinate, and $\nu_0$ is a common angular
velocity. This angular velocity corresponds to the linear velocity $v_0 = \nu_0R$. We
are interested in solutions strongly localized in $\tilde{\varphi}$.

In the case of even $l_0 - N$, the $2\pi$-periodic amplitudes $F(\tilde{\varphi})$ and
$S(\zeta)$ can be expanded in Fourier series
\begin{equation}\label{SS-expansions}
F = \sum\limits_{j} F_j \, e^{{\rm i}j\tilde{\varphi}}, \quad S = \sum\limits_{l} \, S_l \,
e^{{\rm i}l\tilde{\varphi}} \,.
\end{equation}
This means that the frequency spectra of $F$ and $S$ consist of equidistant peaks
separated by the distance $\nu_0 = v_0/R$, i.e., we have a frequency comb. The smaller
the scale of localization of $F(\tilde{\varphi})$ and $S(\tilde{\varphi})$, the broader
is this comb. Note that positions of the equidistant frequency peaks cannot coincide
with non-equidistant modal frequencies. This means that nonlinear frequency shifts must
be involved in formation of the steady states.

In the case of odd $l_0 - N$, when $S(\tilde{\varphi})$ is $2\pi$-periodic and
$F(\tilde{\varphi})$ is $2\pi$-antiperiodic, we must multiply $F$ by the factor of
$\exp({\rm i}\tilde{\varphi}/2)$ to get true $2\pi$-periodic $\tilde{\varphi}$ dependence of
the FH amplitude. This leads us again to a slightly different frequency comb with the
frequency separation~$\nu_0$.

Importantly, velocity $v_0$ (or the angular frequency $\nu_0$) cannot be chosen
arbitrary. It has to be determined simultaneously with nonlinear solution for
$F(\tilde{\varphi})$ and $S(\tilde{\varphi})$. It can be different for the periodic and
antiperiodic solutions.

As follows from Eqs.~(\ref{Initial2},\,\ref{L}), the steady-state amplitudes
$F(\tilde{\varphi})$ and $S(\tilde{\varphi})$ obey the set of nonlinear 
ordinary differential equations
\begin{eqnarray}\label{SS-equations}
&(& \hspace*{-1.5mm} {\rm i}\nu_1^0\partial_{\tilde{\varphi}} + \alpha_1\partial^2_{\tilde{\varphi}} - \Omega_1)F = 2\mu SF^* \\
&(& \hspace*{-1.5mm} {\rm i}\nu_2^0\partial_{\tilde{\varphi}} +
\alpha_2\partial^2_{\tilde{\varphi}} - \Omega_2)S = \mu F^2 + {\rm i}h \,, \nonumber
\end{eqnarray}
where $\nu^0_{1,2} = \nu_{1,2} - \nu_0$. They have to be solved with $2\pi$ periodic and
antiperiodic boundary conditions. The $\nu$ terms account for drift of $F$ and $S$ with
different velocities, while the $\alpha$ terms are relevant to the effects of
dispersion. The limit $\gamma_{1,2} \to 0$, when $\Omega_{1,2} = \Delta_{1,2}$,
corresponds to the absence of dissipation. Generally, nonlinear set~(\ref{SS-equations})
with two complex amplitudes and several variable parameters, including
velocity $v_0$, is much more complicated compared to the single equation describing the
$\chi^{(3)}$ combs.

While we do not know velocity $v_0$ for nonlinear solutions, it is likely that it lies
in between $v_1$ and $v_2$. Setting $|\nu^0_1| \approx |\nu^0_2| \approx |v_1 -
v_2|/2R$, $|\alpha_1| \approx |\alpha_2| \approx |v'|/2R^2$, and assuming that $\delta
\tilde{\varphi} \ll 2\pi$ is the scale of localization of $F(\tilde{\varphi})$ and
$S(\tilde{\varphi})$, we can estimate the ratio of the drift to dispersion terms. It
about $|v_1 - v_2|R \delta\tilde{\varphi}/|v'|$. Adopting parameters of Fig.~\ref{LN-curves}, 
we see that, except for the close vicinity of the point of equal group velocities, we have
$|v_1 - v_2| \gtrsim 10^8$~cm/s and $|v'| \approx 10^4$~cm$^2$/s. For $R \approx 1$~mm,
this gives the ratio $\gtrsim 10^3/\delta \tilde{\varphi}$. Thus, for $\delta
\tilde{\varphi} \gtrsim 10^{-3}$, which means the number of comb peaks $\lesssim 10^3$,
the drift terms are dominating, and the dispersion terms can be omitted in the leading
approximation. In order to omit the drift terms, we must stay practically at the point
$v_1(\lambda_p) = v_2(\lambda_p)$. Slight deviations from this point switch  the strong
drift terms on. This situation is generic for $\chi^{(2)}$ resonators.

Note that the zero point of the frame coordinate $\tilde{\varphi}$ (and of the polar
angle $\varphi$) can be chosen arbitrary. If $f(\tilde{\varphi}),\,s(\tilde{\varphi})$
is a particular steady-state solution, then $f(\tilde{\varphi} -
\varphi_0),\,s(\tilde{\varphi} - \varphi_0)$ with an arbitrary $\varphi_0$ is an equivalent 
solution. This degeneracy in $\varphi_0$ can be crucial for analysis of
corrections to the primary steady states caused by various small perturbations, such as
weak dissipation and dispersion.

\subsection{Dispersionless solitons and periodic states}

Here we demonstrate the possibility of different soliton solutions with the dissipative
and dispersive terms neglected. To do so, we put $\alpha_{1,2} = \gamma_{1,2} = 0$ and
set
\begin{equation}\label{v0}
v_0 = (2\Delta_1v_2 + \Delta_2v_1)/(2\Delta_1 + \Delta_2) \;.
\end{equation}
This common propagation velocity lies inside the interval $[v_1,v_2]$ for
$\Delta_1\Delta_2 > 0$ and outside this interval for $\Delta_1\Delta_2 < 0$.
Furthermore, we change from $\tilde{\varphi}$, $F$ and $S$ to the normalized quantities
$\zeta = R(2\Delta_1 + \Delta_2)\tilde{\varphi}/(v_2 - v_1)$, $f =
\mu(2/|\Delta_1\Delta_2|)^{1/2}F$, and $s = -2\mu S/\Delta_1$. After that we have from
Eqs.~(\ref{SS-equations})
\begin{eqnarray}\label{InitialNormalized}
{\rm i}\dot{u} \hspace*{-1.5mm} &-& \hspace*{-1.5mm} u + s|u| = 0 \\
{\rm i}\dot{s} \hspace*{-1.5mm} &+& \hspace*{-1.5mm} s - q u = \eta \,, \nonumber
\end{eqnarray}
where, as earlier, $\eta = 2|\mu h/\Delta_1\Delta_2|$ is the normalized pump strength, 
$u = f^2$, the dot indicates differentiation in $\zeta$, and $q = {\rm
sign}(\Delta_1\Delta_2) = \pm 1$. The spatially uniform background values of $s$ and $u$
in our case are as follows, compare to Fig.~\ref{Back}: $\bar{s}_{\pm} = \mp 1$,
$\bar{u}_{\pm} = \mp 1 - \eta$ for $q = 1$ ($\Phi = 0$) and $\bar{s} = 1$, $\bar{u} =
\eta - 1$ for $q = -1$ ($\Phi = \pi$).

Remarkably, set (\ref{InitialNormalized}) possesses a family of solutions with real
$s$ and complex $u$. To see it, we indicate that the equality $s - q{\rm Re}(u) = \eta$,
being valid for an arbitrary $\zeta_0$, holds true for any $\zeta$. Transferring next to
the absolute values and arguments of $s$ and $u$, one can make sure that the
conservation law $p_0 \equiv s^2/2 - q|u| = const$ is valid for such solutions. It
corresponds to a constant energy flux along the coordinate $\zeta$. Employing this
conservation law, we obtain a single second-order equation for $s$:
\begin{equation}\label{Single-s}
\ddot{s} = \eta - s(1 + p_0) + s^3/2 \;.
\end{equation}
This equation admits an obvious mechanical analogy with a unit-mass particle with
coordinate $s(\zeta)$ moving in the potential $U(s) = -\eta s + (1 + p_0)s^2/2 - s^4/8$;
$\zeta$ has to be treated as an effective time. For "initial" values $s_0 = s(0)$,
$\dot{s}_0 = \dot{s}(0)$, equation~(\ref{Single-s}) enables one to find the "trajectory"
$s(\zeta)$ for $\zeta \gtrless 0$. Note that the choice of zero point of $\zeta$ is
arbitrary.

The potential profile $U(\zeta)$ is generally different for different trajectories via
the dependence of $p_0$ on $s_0$ and $\dot{s}_0$. The treatment is strongly simplified
if we characterize each trajectory by its turning point $s_0$ corresponding to zero
"velocity" $\dot{s}_0$. In this case, we have $p_0 = s_0^2/2 - q|\eta - s_0|$ according
to Eqs.~(\ref{InitialNormalized}) and, consequently, the potential profile $U =
U(s,s_0)$ in an explicit form:
\begin{equation}\label{U-Explicit}
U = -\eta s + \frac{s^2}{2} \left(1 + \frac{s_0^2}{2} - q|\eta - s_0| \right) -
\frac{s^4}{8} .
\end{equation}
With this profile, one can find, analytically or numerically, trajectory $s(\zeta)$ for
any $s_0$, $\eta$, and $q$.

A key point in our analysis of the localized and periodic solutions for $s(\zeta)$ is
determination and classification of the turning points $s_0$ corresponding to maxima of
$U(s,s_0)$. Equating to zero the derivative $\partial_{s} U(s,s_0)$ at $s = s_0$, we
get: $qs_0|s_0 - \eta| - s_0 + \eta = 0$. Solutions of this equation for $s_0$ depend on
$q = {\rm sign}(\Delta_1\Delta_2)$.
\begin{figure}[h]
\centering 
\includegraphics[width=8cm]{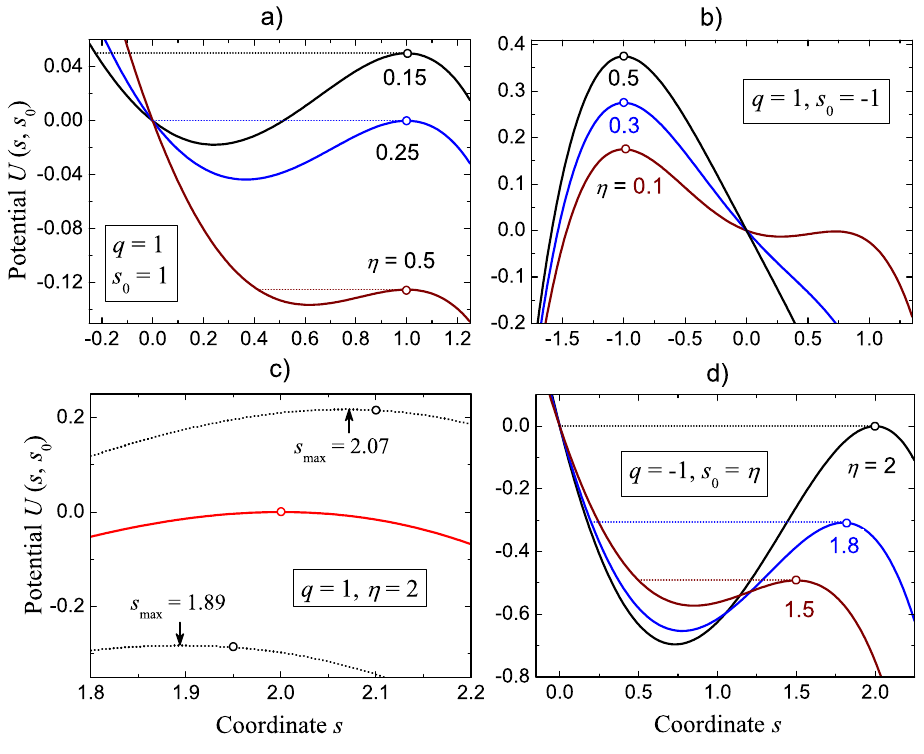}
\caption{Potentials $U(s,s_0)$ for different stopping points $s_0$ (circles) and values
of $\eta$: a) to c) The cases $s_0 = 1$, $-1$, $\eta$ for $q = 1$. Subfigure c) shows
also the effect of small shifts of $s_0$ against $\eta$. d) The case $s_0 = \eta > 1$
for $q = -1$. The horizontal dotted lines correspond to the soliton states.}\label{Pot}
\end{figure}
For $q = 1$, there are three solutions for $s_0$. The first one, $s_0 = \bar{s}_- = 1$,
is valid for $\eta \leq 1$. The corresponding profiles $U(s,1,\eta)$ are illustrated by
Fig.~\ref{Pot}a. The particle can move here between the right (shown by circles) and
left turning points. A small decrease of $s_0$ changes $U(s,s_0)$ such that the particle
oscillates periodically between two turning points, and the oscillation period tends to
infinity for $s_0 \to 1$. For $s_0 > 1$, the regular solution breaks: the particle
infinitely accelerates and $s(\zeta) \to \infty$. The second solution, $s_0 = \bar{s}_+
= -1$, is relevant to the potential profiles of Fig.~\ref{Pot}b. This case is not
physical because solution $s(\zeta)$ breaks for any small deviation of $s_0$ from $-1$.
For the third solution, $s_0 = \eta > 1$, any small deviations of $s_0$ against $\eta$
change the potential such that the particle finds oneself to the right of the potential
maxima, see Fig.~\ref{Pot}c, leading to an infinite growth of $s(\zeta)$. Thus, only the
first solution, where the value $s_0 = 1$ coincides with the background value
$\bar{s}_-$, is physical for $q = 1$.

For $q = 1$, $s_0 = 1$, and $\eta < 1$, Eqs.~(\ref{InitialNormalized}) admit an exact
soliton solution:
\begin{eqnarray}\label{SolitonSolution}
s \hspace*{-1.5mm} &=& \hspace*{-1.5mm} \frac{\sqrt \eta \cosh (\zeta\sqrt{1
\hspace*{-0.5mm} - \hspace*{-0.5mm} \eta}) \hspace*{-0.5mm} + \hspace*{-0.5mm} 2\eta
\hspace*{-0.5mm} - \hspace*{-0.5mm} 1}{\sqrt \eta \cosh (\zeta\sqrt{1
\hspace*{-0.5mm}
- \hspace*{-0.5mm} \eta}) \hspace*{-0.5mm} + \hspace*{-0.5mm} 1}  \\
f \hspace*{-1.5mm} &=& \hspace*{-1.5mm} \pm \sqrt{1 \hspace*{-0.5mm} - \hspace*{-0.5mm}
\eta} \,\frac{\sqrt \eta \sinh (\zeta\sqrt{1 \hspace*{-0.5mm} - \hspace*{-0.5mm}
\eta}) \hspace*{-0.5mm} + \hspace*{-0.5mm} {\rm i}\sqrt{1 \hspace*{-0.5mm} - \hspace*{-0.5mm}
\eta}}{\sqrt \eta \cosh (\zeta \sqrt{1 \hspace*{-0.5mm} - \hspace*{-0.5mm} \eta})
\hspace*{-0.5mm} + \hspace*{-0.5mm} 1} \;. \nonumber
\end{eqnarray}
The normalized SH amplitude $s$ is even in $\zeta$, tends to the background value
$\bar{s}_- = 1$ for $|\zeta| \to \infty$, and possesses the minimum value $2\sqrt{\eta}
- 1$, that can be positive and negative depending on $\eta$. This solution corresponds
to an infinitely long movement of the particle from the right turning point $s_0 = 1$ to
the left turning point and back. The normalized complex FH amplitude $f$ tends to
opposite real values for $\zeta \to \pm \infty$, while $u = f^2$ tends to $u_- = 1 -
\eta$. This means that $f(\zeta)$ is antiperiodic within the interval
$[-\infty,\infty]$. Unless $\eta$ is very close to $1$, approaching the background
values with increasing $|\zeta|$ occurs exponentially, i.e. very fast. For $\eta \to 1$,
the soliton width tends to infinity. The soliton solution
(\ref{SolitonSolution}) has no limit for $\eta \to 0$: For $\eta \ll 1$, the values of
$s(\zeta)$ and $u(\zeta)$ stay very close to the background values $\bar{s}_+ \simeq -1$
and $\bar{u}_+ \simeq -1$ within a finite range of $|\zeta|$ near $0$ and tend
exponentially to $\bar{s}_- \simeq \bar{u}_- \simeq 1$ for larger $|\zeta|$. With
decreasing $\eta$, the left/right boundaries between these background regions shift to
$\pm \infty$.

While the above soliton solution is valid, strictly speaking, only for an
infinite range of $\zeta$, it is of big value for finite ranges, provided that these
ranges substantially exceed the soliton width. First, the difference between the
peripheral values of $s$ and $f$ and the corresponding background values can be smaller
than a natural noise level. Second, existence of soliton solutions ensures, as shown
below, the presence of periodic solutions which are very close to the soliton ones.

\begin{figure}[h]
\centering  
\includegraphics[width=8.2cm]{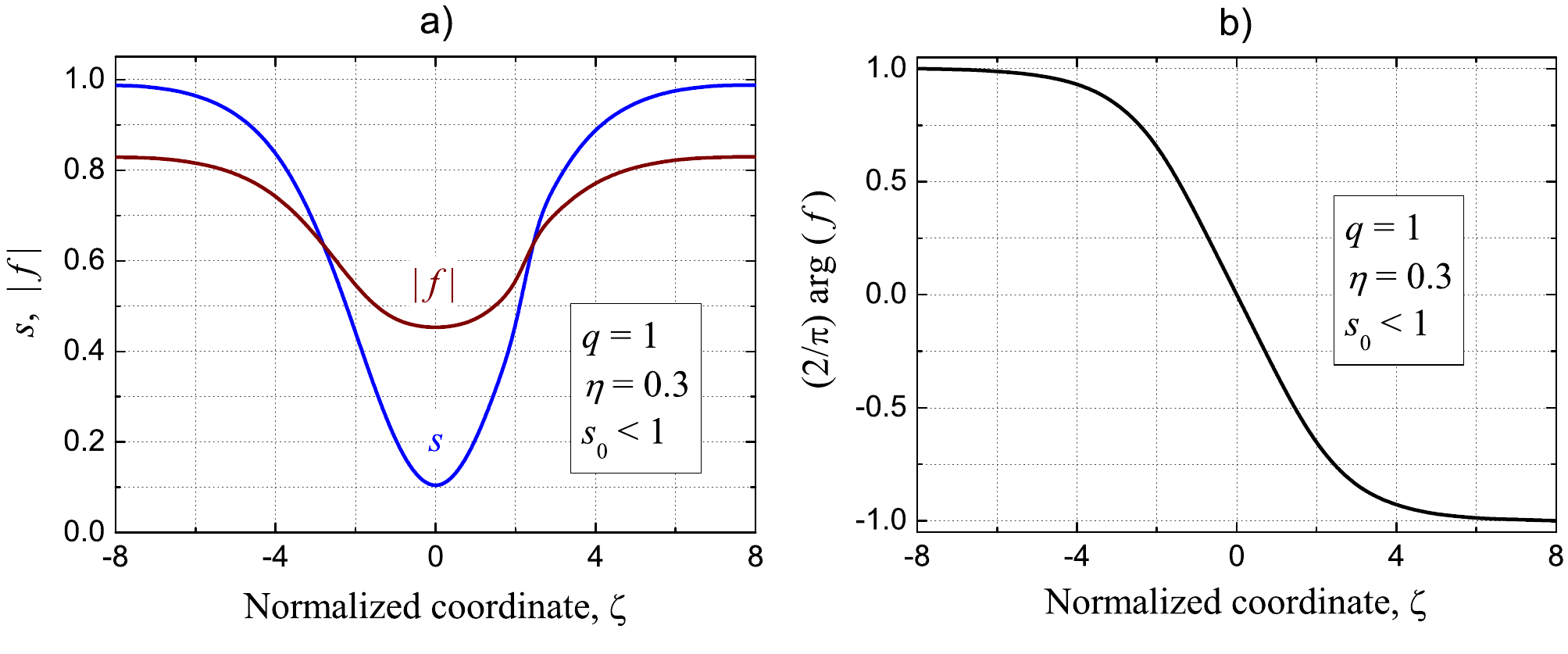}
\caption{Periodic soliton-like solutions for $s$, $|f|$, and ${\rm arg}\,(f)$ for $q =
1$, $\eta = 0.3$, $1 - s_0 \simeq 1.27 \times 10^{-2}$. The period is $\zeta_0 = 16$.
}\label{PeriodicSF}
\end{figure}

To get periodic solutions for $s$ and $f$, we employ the values $s_0$ slightly below
$1$. This is illustrated by Figs.~\ref{PeriodicSF}a,b.
Within the line width, the shape of each dip in a) coincides with the soliton one. The
difference is substantial only when $s(\zeta)$ and $f(\zeta)$ are very close to the
background values $1$ and $\sqrt{1 - \eta} \simeq 0.837$, respectively. The $\pi$-step
dependence of ${\rm arg}\,[f(\zeta)]$ is clearly seen; this shows that our solution
satisfies the antiperiodic boundary conditions. The period $\zeta_0$ depends on the
deviation $\delta s_0 = s_0 - 1$ and tends to infinity for $\delta s_0 \to 1$. This
allows one to adjust the period to the dimensionless resonator length $\zeta_R = 2\pi
R|2\Delta_1 + \Delta_2|/|v_1 - v_2|$. For sufficiently large values of $\zeta_R$,
periodic multi-soliton solutions become possible.

Now we switch to the case $q = {\rm sign}(\Delta_1\Delta_2) = -1$. Only the values $s_0
= \eta$ for $\eta > 1$, which correspond to $p_0 = \eta^2/2$, give here potential
profiles with maxima at $s = s_0$. This is illustrated by Fig.~\ref{Pot}c. For each
value $s_0 = \eta$, there is a left partner turning point providing the soliton state.
In contrast to the case $q = 1$, $s_0 = \eta$, a small deviation $\delta s_0 = s_0 -
\eta$ of any sign changes the potential $U(s,s_0)$ such that $s_0$ stays on the left of
the potential maximum. In other words, there is a family of periodic states approaching
the soliton state for $\delta s_0 \to 0$ regardless of the sign of $\delta s_0$. Note
that the point $s_0 = \bar{s} = 1$, relevant to the spatially uniform background,
corresponds to a minimum of $U(s,s_0)$; it is of no interest.

For $q = -1$, $s_0 = \eta$, and $\eta > 1$, Eqs.~(\ref{InitialNormalized}) also admit an
exact soliton solution:
\begin{eqnarray}\label{Soliton2}
s \hspace*{-1mm} &=& \hspace*{-1mm} \frac{\eta \cosh (a\zeta ) - \eta^2 + 2}{\cosh (a\zeta) + \eta } \\
f \hspace*{-1mm} &=& \hspace*{-1mm} \pm \sqrt{2} a\, \frac{\sqrt{\eta \hspace*{-0.5mm} +
\hspace*{-0.5mm} 1}\cosh (a\zeta/2) \hspace*{-0.5mm} - \hspace*{-0.5mm}
{\rm i}\sqrt{\eta \hspace*{-0.5mm} - \hspace*{-0.5mm} 1}\sinh (a\zeta/2)}{\cosh (a\zeta) + \eta}, \nonumber
\end{eqnarray}
where $a = \sqrt{\eta^2 - 1}$. It corresponds to $s(0) = 2 - \eta < 1$, $s(\pm \infty) =
\eta$, $f(0) = \pm \sqrt{2(\eta - 1)}$, and $f(\pm \infty) = 0$. This soliton is thus
SH-dark and FH-bright. Approaching the limiting values at $\pm \infty$ occurs
exponentially. The localization scale substantially decreases with increasing $\eta$.
Furthermore, the argument of $f$ is an odd function of $\zeta$, whose asymptotic values
at $\pm \infty$ are below $\pi/4$ in the absolute value. Note also a useful link  
$|f|^2 = (\eta^2 - s^2)/2$. Further details on this soliton are given below in co-junction 
with an analysis of closely related soliton-like periodic states.

For $|\delta s_0| \ll 1$, we have a family of periodic soliton-like solutions $s(\zeta)$
and $f(\zeta)$; they can be found numerically from Eqs.~(\ref{InitialNormalized}) or
(\ref{Single-s}). The cases $\delta s_0 \lessgtr 0$ have not only similarities, but also
important special features.
\begin{figure}[h]
\centering  
\includegraphics[width=8.4cm]{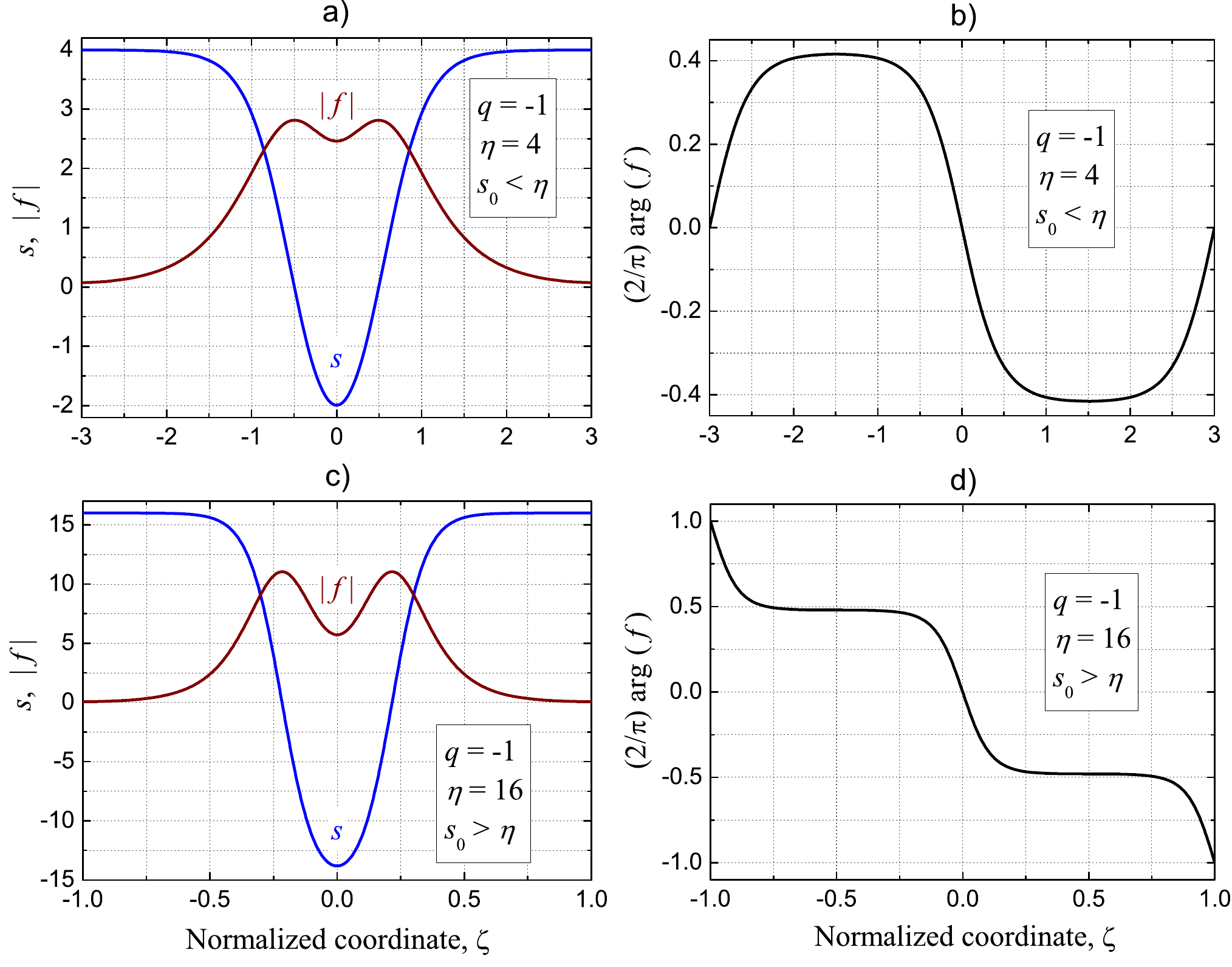}
\caption{Periodic soliton-like solutions for $s$, $|f|$, and ${\rm arg}\,(f)$ for $q =
-1$ within one period. Subfigures a) and b) correspond to $\eta = 4$, $1 - s_0 \simeq
0.0035$, and to the period $\zeta_0 = 6$, while subfigures c) and d) refer to $\eta =
16$, $1 - s_0 \simeq 0.005$, and $\zeta_0 = 2$. Note the difference in the horizontal
and vertical scales between a,b and c,d. }\label{PeriodicFour}
\end{figure}
Figures~\ref{PeriodicFour}a,b illustrate the case $\delta s_0 < 0$ for $\eta = 4$. Here
the period is $\zeta_0 = 6$. Within the line width, the curves $s(\zeta)$ and
$|f(\zeta)|$ coincide with ones given by Eqs.~(\ref{Soliton2}). The function
$|f(\zeta)|$ has a characteristic two-hump structure. The tails ($|\zeta| > 2$), where
the difference with the soliton is relatively large, are weak. In contrast to the case
$q = 1$, the function ${\rm arg}[f(\zeta)]$ is $\zeta_0$-periodic, it experiences modest
oscillations with $\zeta$.

Figures~\ref{PeriodicFour}c,d illustrate the case $\delta s_0 > 0$ for $\eta = 16$. Now,
the period of $s(\zeta)$ $|f(\zeta)|$ is $\zeta_0 = 2$, the vertical scale in c) is
enlarged compared to a), and the tails of $s(\zeta)$ and $|f(\zeta)|$ are weakened. This
effect of increasing $\eta$ is relevant also to the case $\delta s_0 < 0$. The
difference is in the behavior of ${\rm arg}[f(\zeta)]$, as shown in d). This function is
now step-like, so that the function $f(\zeta)$ is $\zeta_0$-antiperiodic.

Regardless of the sign of $\delta s_0$, increasing $|\delta s_0|$ results in decreasing
period $\zeta_0$, broadening of the deeps/peaks of $s({\zeta})$ and $|f(\zeta)|$, and in
strengthening of the tails. The values of $\delta s_0$ are not necessarily small. The
effect of increasing $\eta$ is different. It is decreasing period ($\zeta_0 \propto
1/\eta$ for $\eta \gg 1$), narrowing of the deeps and peaks, and suppression of the
tails. In other words, the periodic state verges towards a periodic train of solitons
with increasing $\eta$.  The case of almost opposite $\Delta_2$ and $2\Delta_1$ is
unfavorable for solitons.

Requirements to the experimental parameters $R$, $\Delta_{1,2}$, and $\eta$, necessary
to realize periodic soliton-like solutions, are different for the cases $q = 1$ and
$-1$. In the case $q = 1$, when $\eta < 1$, the period $\zeta_0$ cannot be smaller than
$1$, but is often $\sim 10$. Therefore, the dimension resonator length $\zeta_R = 2\pi
R|2\Delta_1 + \Delta_2|/|v_1 - v_2|$ has to be much larger than $1$. For $2\pi R \approx
1$~cm and $|\Delta_1| \approx |\Delta_2|$, this leads us to detunings $|\Delta_{1,2}|
\gtrsim 10^9$~s$^{-1}$. The restriction from above $|\Delta_{1,2}| \ll \delta \omega  =
v_{1,2}/R \approx 10^{11}$~s$^{-1}$ can be easily fulfilled. In the case $q = -1$, when
$\eta$ can be much larger than $1$, the period $\zeta_0$ can be substantially smaller
than~$1$. Here the the detunings $|\Delta_{1,2}|$ can stay below $10^8$~s$^{-1}$.
Keeping in mind that the decay constants $\gamma_{1,2}$ must be much smaller than
$|\Delta_{1,2}|$, we see that substantially larger modal quality factors can be used.

The above theoretical treatments are relevant to the case $\gamma_{1,2} =v'_{1,2} = 0$. 
It is possible to get small corrections $\delta s(\tilde{\varphi})$ and $\delta f(\tilde{\varphi})$ 
to the above steady-state solutions within the linear approximation in small parameters $\gamma_2$ 
and $v'_{1,2}$. Such a perturbation approach~\cite{ZeroModes} accounts for the degeneracy in the 
choice of position of the soliton center and is consistent with the known Fredholm alternative for 
linear systems. However, the linear perturbation theory in $\gamma_1$ fails indicating a substantial 
modification of the initial soliton state. This means that more powerful perturbation techniques, 
like the Newton method, see~\cite{Barashenkov96} and references therein, have to be employed. 
Such a situation is not rare in the theory of solitons. In our case, a strong effect of 
$\gamma_1$ on solitons is closely related to the strong effect on the spatially uniform background, 
see Fig.~\ref{Background}. A similar situation is known for Kerr solitons~\cite{Barashenkov96}. 
Consideration of the impact of $\gamma_1$ on the soliton properties requires a separate study.   

\section{Numerical simulations}

We investigated numerically stability of the found steady-state solutions within the
modal approach relevant to ordinary differential equations~(\ref{Initial}) and,
independently, within the slowly-varying amplitude approach relevant to partial
differential equations~(\ref{Initial2}). The modal amplitudes $F_j(t)$ and $S_l(t)$ are
linked to the $2\pi$-periodic amplitudes $F(\varphi,t)$ and $S(\varphi,t)$ by the
discrete Fourier transform. Both Eqs.~(\ref{Initial}) and~(\ref{Initial2}) were solved
in time domain. As an initial condition for $F_j,S_l$, we used the inverse Fourier
transform of a periodic soliton-like solution $F(\tilde{\varphi})$, $S(\tilde{\varphi})$
superimposed by very small random numbers $\delta F_j,\delta S_l$ (modal noise). As
initial conditions for $F(\varphi,t)$ and $S(\varphi,t)$, we employed
$F(\tilde{\varphi})$, $S(\tilde{\varphi})$ superimposed by the Fourier transform of the
modal noise. Different variants of periodic and antiperiodic initial conditions with
$\Delta_1\Delta_2 \gtrless 0$ were used.

The number of points in the $\varphi$ mesh and, correspondingly, the number of modes
within each of the FH and SH frequency domains ranged from $512$ to $2048$. The
nonlinear set of ordinary differential equations~(\ref{Initial}) was integrated using
fourth-order Runge-Kutta method with the time step ranging from $5 \times 10^{-4}$ to $2
\times 10^{-3}$ of the round trip time $R/c$ with $R = 1.59$~mm. The nonlinear partial
differential equations for $F(\varphi,t)$ and $S(\varphi,t)$ were solved using the
step-split Fourier method~\cite{ssfm}, the step in time ranged from $5 \times 10^{-6}$
to $2 \times 10^{-5}$ of $R/c$. It was made sure that both methods give essentially the
same results and no real effect of time- and $\varphi$-steps is present. Also numerical
solutions were checked in a series of limiting cases of initial conditions where the
analytical solutions can be obtained easily.

\begin{figure}[h]
\centering
\includegraphics[width=8.4cm,height=8.5cm]{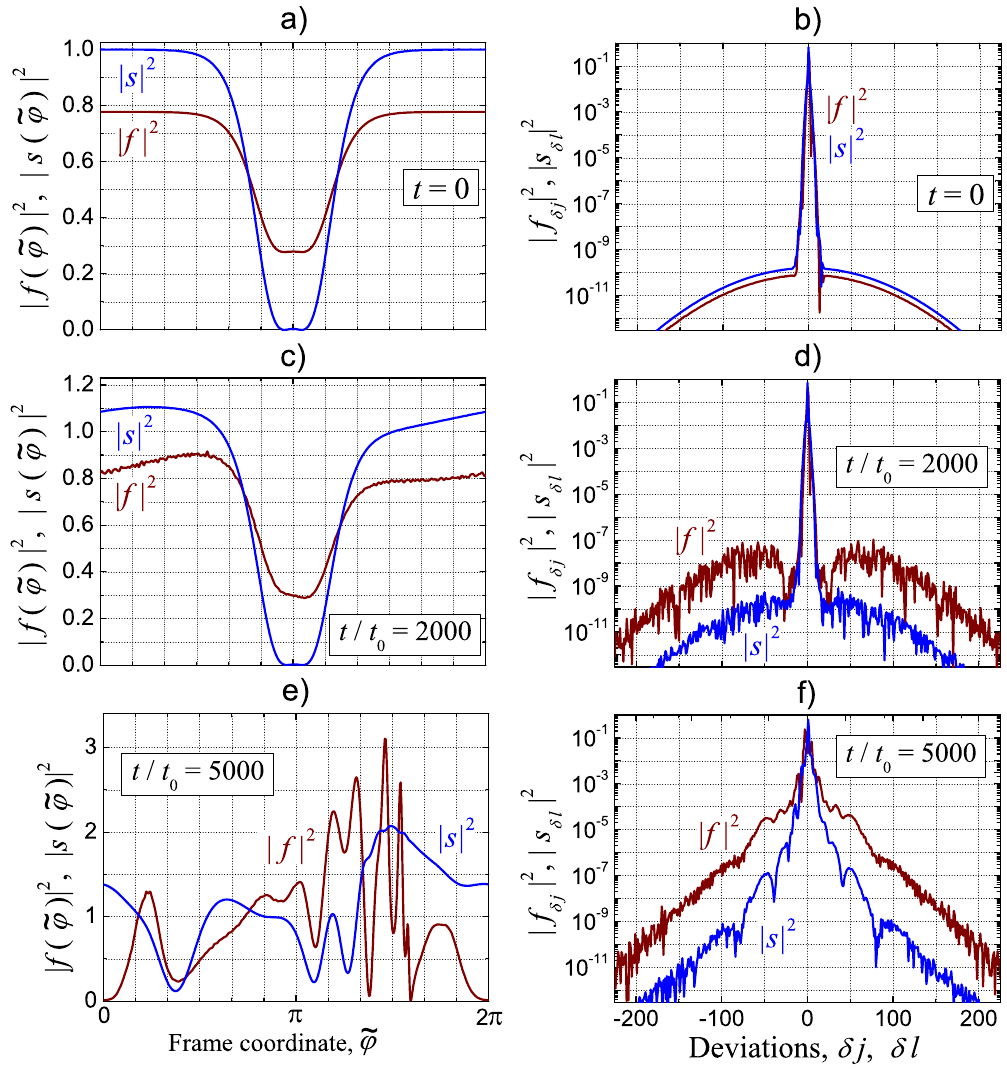}
\caption{Numerical results on stability of the antiperiodic soliton-like solution with
$q = 1$ and $\eta = 0.22$ ($2\pi R = 1$~cm, $\Delta_1 = \Delta_2 = 3 \times
10^9$~s$^{-1}$, $v_1 - v_2 = 10^{-2}c$, $\gamma_{1,2} = 0$, and $v'_{1,2} = 0$). The
time $t$ is normalized to $t_0 = R/c \simeq 5 \times 10^{-12}$~s. The left column shows
the normalized intensities $|f|^2$ and $|s|^2$ versus $\varphi$ and the right column
shows the normalized modal powers $|f_{\delta j}|^2$ and $|s_{\delta j}|^2$ versus the
deviations of the modal numbers $\delta j = j - j_0$ and $\delta l = l - l_0$.
}\label{Numericals}
\end{figure}
Figure~\ref{Numericals} exhibits representative results of our numerical simulations for
an antiperiodic soliton-like solution relevant to $q = {\rm sign}(\Delta_1\Delta_2) = 1$
and $\eta = 0.22$. Initially ($t < 10^{-8}\,{\rm s}$), the noise of $f_{\delta j}$ grows
exponentially with an increment $\approx \Delta_{1,2}$ within a wide range of deviations
$|\delta j|$, see subfigures b) and d). The normalized intensity dependences
$|f|^2(\tilde{\varphi})$ and $|s|^2(\tilde{\varphi})$ experience only minor changes at
this stage, and the FH and SH intensity patterns move practically with the same velocity
$v_0$, see a) and b). For larger times, distortions of the steady-state solution become
substantial, the notion of a common velocity fails, and a strong fragmentation of both
patterns in $\tilde{\varphi}$ takes place, see e). At the same time, both patterns
remain coherent. The spectral broadening is continuing, see d).

Changes of the simulation and external parameters, switching from the antiperiodic to
periodic solutions and to the case $q = -1$, inclusion of nonzero $\gamma_{1,2}$ and
$v'_{1,2}$ results in modification of details, but the fact of instability remains
unchanged. At the same time, solutions with a zero noise stay unchanged indefinitely
long; this confirms correctness of the simulations.

Finally, we stress that the above numerical analysis refers to the soliton 
solutions calculated at $\gamma_{1,2} = 0$ and $v^0_{1,2} = 0$. A strong impact of 
$\gamma_1$ on the soliton shape can bring the system to stability by analogy with 
Kerr solitons~\cite{Barashenkov96}.

\section{Discussion}

While the differences between our $\chi^{(2)}$ case and the case of $\chi^{(3)}$ combs
are quite evident, distinctions from the known analyzes of $\chi^{(2)}$ solitons have to
be considered. The former results were relevant mostly to the conservative case (zero
pumping and dissipation) and infinite media~\cite{SkryabinReview}. A considerable part
of theoretical considerations was relevant to the spatial solitons, which are essentially 
different from the temporal solitons. Applications to microresonators, including 
discreteness of the modal structure, practicability of the chosen parameters, and the 
possibilities of pumping were not considered. Three recent papers on 
dissipative solitons in microresonators~\cite{Wabnitz18,Skryabin19,Skryabin19A} are 
relevant to FH pumping at the point of equal group velocities, $v_1(\lambda_p) = v_2(\lambda_p)$.

Each of the general assumptions -- single-mode approximation and perfect radial poling
-- is rather obligatory. Often, there are plenty of competing modes characterized not
only by the azimuth number, but also the radial and polar
numbers~\cite{Vahala03,ModalStructure,Identification}. It is unlikely that account for
these modes can be compatible with analysis of coherent soliton-like states. Suppression
of undesirable modes in high-$Q$ resonators is a serious technological task that requires 
special efforts~\cite{ModeManagement1,ModeManagement2,ModeManagement3}. Radial poling is 
generally imperfect because of off-centering of the domain structure~\cite{RadialPoling1}. 
This leads to the presence a few (or several) Fourier peaks of the quadratic susceptibility 
coefficient, separated by the intermodal distance, instead of a single $N$-peak in Fig.~\ref{Geometry}d. 
Unless only one of these peaks is dominating, equations for slowly varying amplitudes $F(\varphi,t)$ 
and $S(\varphi,t)$ can fail. The use of the linear poling~\cite{IlchenkoPRL04}, leading to broad 
Fourier spectra of the quadratic susceptibility, is, most probably, very harmful for coherent
solutions.

A substantial difference of the group velocities for the first and second harmonics is a
generic feature of $\chi^{(2)}$ microresonators. For many resonators, the difference
$|v_1 - v_2|$ can be estimated as $\sim 10^{-2}c = 3 \times 10^{8}$~cm/s within broad
spectral ranges of the pump wavelength $\lambda_p$. Any steady-state comb solution
$F(\varphi - \nu_0t)$, $S(\varphi - \nu_0t)$ implies that the difference between $v_1$
and $v_2$ is compensated by the quadratic nonlinearity. This compensation is
fundamentally different from the known compensation between dispersion and nonlinearity
for cubic solitons. Determination of coherent steady-state solutions with nonlinearly
compensated velocity difference is by itself a big challenge. Above we have found two
big families of such localized soliton solutions. The effects of dispersion can be
estimated here as secondary for not extremely narrow FH/SH solitons. Specifically, they
are weak until the comb spectrum consists of $\sim 10^3$ (or less) frequency peaks.

Indeed, the velocity difference can be suppressed when working at the point
$v_1(\lambda_p) = v_2(\lambda_p)$. In this case, the situation becomes similar to that
typical of $\chi^{(3)}$ combs. However, the question about the effect of inevitably
present relatively small velocity differences remains open.

A surprising feature of the found nonlinear regimes is the presence of two entirely
different types of spatial symmetry in the case of SH pumping, $2\pi$-periodic and
-antiperiodic solutions. While the total electromagnetic fields are ultimately
$2\pi$-periodic, these solutions are topologically different. In particular, the
antiperiodic solutions have no spatially uniform backgrounds. Both periodic and
antiperiodic solitons admit simple analytical representations which are new to the best
of our knowledge. Furthermore, each soliton is associated with a family of soliton-like
periodic states with adjustable spatial periods.

Unfortunately, our numerical simulations have shown that the found dispersionless
solitons and periodic states are not stable against the growth of small perturbations
leading to broadening of the frequency spectrum. At the same time, the spatial coherence
is still there. Most probably, this means that the model with omitted weak frequency
dispersion is oversimplified. Not only the group velocity difference, but also
dispersion broadening, have to be simultaneously compensated by nonlinearity. Further
theoretical efforts are necessary to establish whether $\chi^{(2)}$ frequency combs
relevant to solitons are achievable.

\section{Conclusions}

We have derived and analyzed coupled nonlinear equations for FH and SH envelopes
relevant to frequency comb generation in $\chi^{(2)}$ microresonators. They incorporate
selective SH pumping, quasi-phase-matching via radial poling, different group velocities
and frequency dispersions, and modal decay. Within a broad range of parameters, the
effect of the group velocity difference is found to be crucial for comb related
solutions. It is shown that, depending on the pump frequency, solutions for the modal
envelopes can be topologically different and relevant to periodic and antiperiodic
boundary conditions. Within a model disregarding weak dispersion, we have found
analytically and numerically two families of dissipative steady-state soliton and
soliton-like solutions. These solutions correspond to the FH and SH envelopes moving
with the same velocity and possessing equidistant frequency spectra, they can be
periodic and antiperiodic. Numerical calculations show, nevertheless, that weak
dispersion terms are indispensable in providing stability of comb solutions.

\noindent {\bf Acknowledgement:} The work of S.S. was supported by the Russian
Foundation for Basic Research (project 18-29-20025).


\end{document}